\documentclass[10pt,twocolumn,letterpaper]{article}

\usepackage[pagenumbers]{cvpr} 

\usepackage{multirow}

\usepackage[ruled, linesnumbered]{algorithm2e} 

\SetKwInput{KwInput}{Input}                
\SetKwInput{KwOutput}{Output}              
\SetAlFnt{\small}
\SetAlCapFnt{\small}
\SetAlCapNameFnt{\small}
\SetAlCapHSkip{0pt}

\usepackage{amsfonts}
\usepackage{amsmath}
\newcommand{\R}{\mathbb{R}}
\newcommand{\grad}[1]{\nabla#1}
\newcommand{\jac}[1]{\mbox{\textbf{J}} #1}
\newcommand{\cdott}{\boldsymbol{\cdot}}

\newcommand{\dist}[2]{\text{d}\left(#1,#2\right)}
\newcommand{\elod}{\delta}
\newcommand{\abs}[1]{\big|#1\big|}

\newtheorem{theorem}{Theorem}
\newtheorem{proposition}[theorem]{Prop.}

\makeatletter
\@namedef{ver@everyshi.sty}{}
\makeatother
\usepackage{todonotes}
\usepackage[inline]{enumitem}

\usepackage{soul,color}

\usepackage[numbers,sort,compress]{natbib}

\usepackage[pagebackref,breaklinks,colorlinks]{hyperref}

\usepackage[capitalize]{cleveref}
\crefname{section}{Sec.}{Secs.}
\Crefname{section}{Section}{Sections}
\Crefname{table}{Table}{Tables}
\crefname{table}{Tab.}{Tabs.}

\def\cvprPaperID{3966} 
\def\confName{CVPR}
\def\confYear{2023}

\begin{document}


\title{Neural Implicit Mapping via Nested Neighborhoods}

\author{
\normalsize Vinicius da Silva\\
\small PUC-Rio\\
\and
\normalsize Tiago Novello\\
\small IMPA
\and
\normalsize Guilherme Schardong\\
\small U Coimbra
\and
\normalsize Luiz Schirmer\\
\small U Coimbra
\and
\normalsize Helio Lopes\\
\small PUC-Rio
\and
\normalsize Luiz Velho\\
\small IMPA
}

\maketitle

\begin{abstract}
We introduce a novel approach for rendering \textbf{static and dynamic} 3D neural signed distance functions (SDF) in real-time. We rely on nested neighborhoods of zero-level sets of neural SDFs, and mappings between them. This framework \textbf{supports animations} and achieves real-time performance \textbf{without the use of spatial data-structures}. It consists of three uncoupled algorithms representing the rendering steps. The \textbf{multiscale sphere tracing} focuses on minimizing iteration time by using coarse approximations on earlier iterations. The \textbf{neural normal mapping} transfers details from a fine neural SDF to a surface nested on a neighborhood of its zero-level set. It is smooth and it does not depend on surface parametrizations. As a result, it can be used to \textbf{fetch smooth normals for discrete surfaces} such as meshes and to skip later iterations when sphere tracing level sets. Finally, we propose an algorithm for \textbf{analytic normal calculation for MLPs} and describe ways to obtain sequences of neural SDFs to use with the algorithms.



\end{abstract}


\section{Introduction}

\textit{Neural signed distance functions} (SDF) emerged as an important model representation in computer graphics. They are neural networks representing SDF of surfaces, which can be rendered by finding their zero-level sets. The \textit{sphere tracing} (ST)~\cite{hart1989ray, hart1996sphere} is the standard algorithm for this~task.

Neural SDFs created new paths of research challenges, including how to render their level sets flexibly and efficiently. In this work, we approach the problem of \textit{real-time rendering} of neural SDFs level sets. Previous approaches train specific spatial data-structures, which may be restrictive for dynamic models. Another option is to extract the level sets using marching cubes, which needs precomputation, results in meshes that require substantially more storage than neural SDFs, and depends on grid resolution.

We propose a novel framework for real-time rendering of neural SDFs based on \textit{nested neighborhoods} and mappings between them. It supports animations and achieves real-time performance without the use of spatial data-structures. We define three algorithms to be used with such neural SDFs. The \textit{multiscale ST} uses the nested neighborhoods to minimize the overhead of iterations, relying on coarse versions of the SDF for acceleration. 

Given a coarse surface $S$ nested in a neighborhood of the zero-level set of a detailed neural SDF, the \textit{neural normal mapping} transfers the detailed normals of the level sets to $S$. This procedure is smooth and is very appropriate to transfer details during the rendering because it does not depend on parametrizations. 
As a result, it can be used to fetch smooth normals for discrete surfaces such as meshes and to skip later iterations when sphere tracing level sets.

Finally, we propose an algorithm for analytic normal calculation for MLPs and describe ways to obtain sequences of neural SDFs to use with the algorithms.
Summarizing, the contributions of this framework are the following:

\begin{itemize}
    \item Real-time rendering of static and animated 3D neural SDFs without spatial data-structures;
    \item Multiscale ST to minimize ST iteration overhead;
    \item Neural normal mapping to transfer the level set normals of a finer neural SDF to the level sets of a coarser~SDF or to a discrete surface; 
    \item Analytic normal calculation for MLPs to compute smooth normals;
    \item Methods to build sequences of neural SDFs with the neighborhoods of their zero-level sets nested.
\end{itemize}

\section{Related Work}
\label{s-related_works}

Implicit functions are an essential topic in
computer graphics ~\cite{velho2007implicit}.
SDFs are important examples of such functions~\cite{bloomenthal1990interactive} and arise from solving the Eikonal problem~\cite{sethian2000fast}.
Recently, neural networks have been used to represent SDFs~\cite{park2019deepsdf,gropp2020implicit,sitzmann2020implicit}. \textit{Sinusoidal networks} are an important example, which are \textit{multilayer perceptrons} (MLPs) having the sine as its activation function. 
We use the framework in~\cite{novello21diff} to train the networks which consider the parameter initialization given in~\cite{sitzmann2020implicit}.

\textit{Marching cubes} \cite{lorensen1987marching,lewiner2003efficient} and \textit{sphere tracing}~\cite{hart1989ray,hart1996sphere} are classical visualization methods for rendering SDF level sets. 
Neural versions of those algorithms can also be found in~\cite{liao2018deep,chen2021neural,liu2020dist}.
While the initial works in neural SDFs use marching cubes to generate visualizations of the resulting level sets~\cite{gropp2020implicit,sitzmann2020implicit, park2019deepsdf}, recent ones have been focusing on sphere tracing, since no intermediary representation is needed for rendering~\cite{davies2020overfit,takikawa21nglod}. We take the same~path.

Fast inference is needed to sphere trace neural SDFs.
\citet{davies2020overfit} shows that this is possible using General Matrix Multiply (GEMM) \cite{dongarra1990set,tiny-cuda-nn}, but the capacity of the networks used there does not seem to represent geometric detail. Instead, we use the neural SDF framework in~\cite{novello21diff} to represent detailed geometry.
Our approach is to show that ST can be adapted to handle multiple neural SDFs and that we can transfer details from them during the process.
We also compute the analytic neural SDF gradients to improve shading performance and accuracy during rendering.

State-of-the-art works in neural SDFs store features in the nodes of \textit{octrees}~\cite{takikawa21nglod, martel21acorn}, or limit the frequency band in training~\cite{lindell2021bacon}. 
Octree-based approaches cannot directly handle dynamic models. On the other hand, our method does not need spatial data structures, supports animated surfaces, and provides smooth normals which are desired during~shading.

\textit{Normal mapping}~\cite{cohen1998appearance,cignoni1998general} is a classic method to transfer detailed normals between meshes, inspired by \textit{bump mapping}~\cite{blinn1978simulation} and \textit{displacement mapping}~\cite{krishnamurthy1996fitting}. 
Besides depending on interpolation, normal mapping also suffers distortions of the parametrization between the underlying meshes, which are assumed to have the same topology.
Using the continuous properties of neural SDFs allows us to map the gradient of a finer neural SDF to a coarser one. This mapping considers a volumetric neighborhood of the coarse surface instead of parametrizations and does not rely on interpolations like the classic one. 

\section{Nested Neighborhoods of neural SDFs}

This section describes our method. The objective is to render level sets of neural SDFs in real-time and in a flexible way. Given the iterative nature of the ST, a reasonable way to increase its performance is to optimize each iteration or avoid them. The key idea is to consider neural SDFs with a small number of parameters as an approximation of earlier iterations and map the normals of the desired neural SDF to avoid later iterations. We show below that both approaches can be made by mapping neural SDFs using nested neighborhoods. 

The basic idea comes from the following fact: if the zero-level set of a neural SDF $f$ is contained in a neighborhood $V$ of the zero-level set of another neural SDF, then we can map $f$ into $V$. 
This approach results in novel algorithms for sphere tracing, and normal mapping.

\subsection{Overview}

\begin{figure}[hh]
\centering
    \includegraphics[width=0.8\columnwidth]{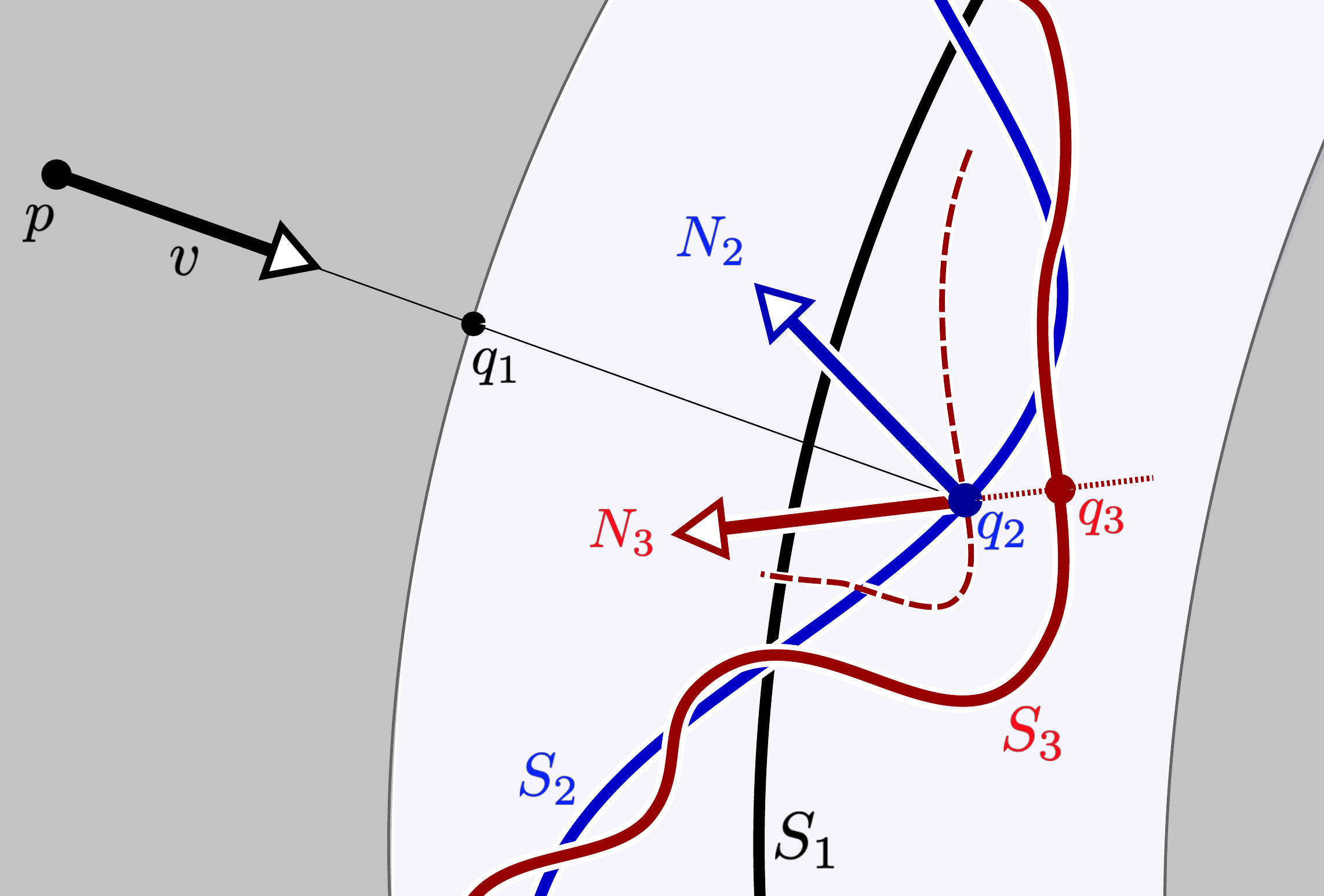}
    \caption{Overview. Let $S_1$, $S_2$, $S_3$ be a sequence of surfaces pairwise close. 
    \textit{Multiscale ST}:
    to sphere trace $S_2$ we first sphere trace the boundary of a neighborhood of $S_1$ (gray), resulting in $q_1$. Then we continue to sphere trace $S_2$, reaching $q_2$.
    \textit{Neural normal mapping}:
    since $q_2$ belongs to a (tubular) neighborhood of $S_3$, we evaluate the normal $N_3$ at $q_2$ of a parallel surface of $S_3$ (red dotted). Notice that these surfaces share the same normal field.}
    \label{f-overview}
\end{figure}

We focus on a simple example with three SDFs as an overview. We follow the notation in Fig.~\ref{f-overview}.
Let $S_1$, $S_2$, $S_3$ be surfaces pairwise close. We assume their SDFs $f_1$, $f_2$, $f_3$ to be sorted by complexity, so evaluating $f_1$ is easier than $f_2$ and $f_3$. 
We use $S_1$ and $S_2$ to illustrate the multiscale ST and $S_3$ to illustrate the neural normal mapping.

\subsubsection*{{Multiscale ST}} Suppose that the ray $p+tv$, with origin at a point $p$ and direction $v$, intersects $S_2$ at a point $q_2$. 
To compute $q_2$, we first sphere trace the boundary of a neighborhood of $S_1$ (gray) containing $S_2$, by using $f_1$. This results in $q_1$. 
Then we continue to sphere trace $S_2$ using $f_2$, reaching $q_2$.
In other words, we are mapping the values of $f_2$ to the neighborhood of $S_1$.
Using an inductive argument allows us to extend this idea to a sequence of SDFs. See Sec. \ref{s-definition} for details.

\subsubsection*{{Neural Normal Mapping}}
For shading purposes, we need a normal vector at $q_2$. 
This can be achieved by evaluating the gradient $N_2=\nabla f_2(q_2)$ of $f_2$ at $q_2$. Instead, we propose to pull the finer details of $S_3$ to $S_2$ to increase fidelity. This is done by mapping the normals from $S_3$ to $S_2$ using $N_3=\nabla f_3(q_2).$

To justify this choice observe that $q_2$ belongs to a (tubular) neighborhood of $S_3$. Thus, $N_3$ is the normal of $S_3$ at its closest point $q_3=q_2-\epsilon N_3$, where $\epsilon$ is the distance from $q_2$ to $S_3$ given by $f_3(q_2)$.
Thus, we are transferring the normal $N_3$ of $S_3$ at $q_3$ to $q_2$.
Observe that $N_3$ is also the normal of the $\epsilon$-level set of $f_3$ at $q_2$ (red dotted).

\subsection{Definition}
\label{s-definition}
A \textit{neural SDF} $f_\theta:\R^3\to \R$ is a smooth neural network with parameters $\theta$ such that $|\grad{f_\theta}|\approx 1$. 
We call its \textit{zero-level set} a \textit{neural surface} and denote it by $S_\theta$.
 
Let $f_{\theta_1}$, $f_{\theta_2}$ be neural SDFs.
We say that $f_{\theta_2}$ is \textit{nested} in $f_{\theta_1}$ for \textit{thresholds} $\elod_1,\elod_{2}>0$ if the $\elod_2$-neighborhood of $S_{\theta_2}$ is \textit{contained} in the $\elod_1$-neighborhood of $S_{\theta_1}$:
\begin{equation}\label{e-LOD}
    \big[|f_{\theta_2}|\leq\elod_2\big]\subset \big[|f_{\theta_{1}}|\leq\elod_{1}\big].
\end{equation}
For simplicity, we denote it by $f_{\theta_1}\rhd f_{\theta_{2}}$ and omit the thresholds $\elod_i$ in the notation.
See Sec.~\ref{s-examples} for examples.

\begin{figure}[hh]
\centering
    \includegraphics[width=0.8\columnwidth]{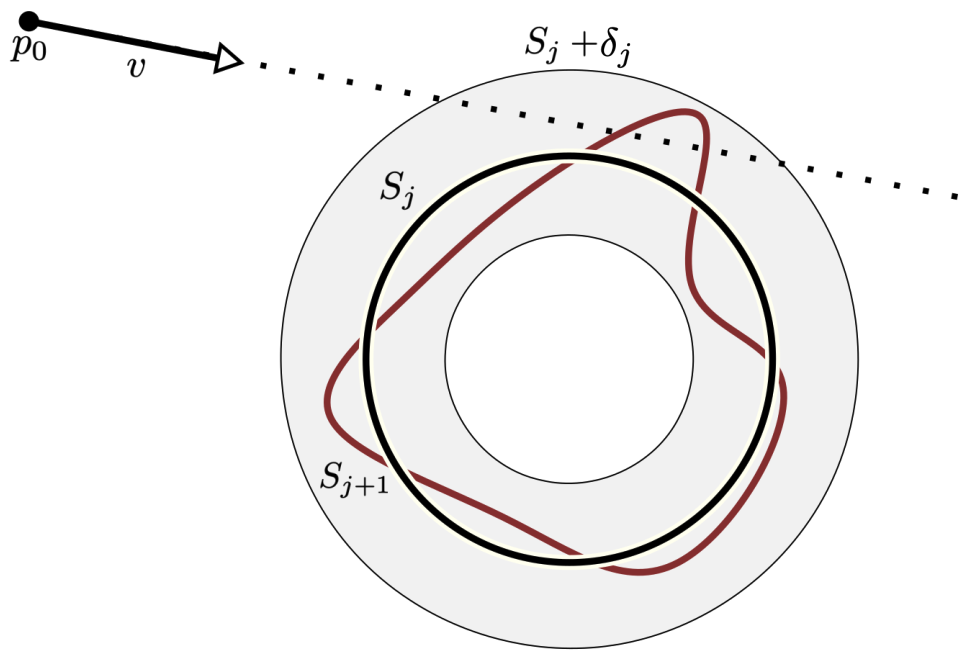}
    \caption{Illustration of a ray intersecting a pair of neural surfaces $S_{j}$ and $S_{j+1}$ with their neural SDFs satisfying $f_{\theta_j}\rhd f_{\theta_{j+1}}$.
    }\label{f-sphere_tracing}
\end{figure}

A sequence of neural SDFs $f_{\theta_1}, \ldots, f_{\theta_m}$ is \textit{nested} if $f_{\theta_{j}}\rhd f_{\theta_{j+1}}$ for $j=1,\ldots,m-1$.
We use $S_j$ to denote their neural surfaces.
The nesting condition implies that each neural surface $S_{j+1}$ is contained in $\big[|f_{\theta_{j}}|\leq\elod_{j}\big]$. 
Thus, to sphere trace $S_{j+1}$ we can first sphere trace the $\delta_j$-level set $S_{j}+\elod_j$ of $f_{\theta_j}$, then continue using $f_{\theta_{j+1}}$ (see Fig.~\ref{f-sphere_tracing}).

We can also extend these sequences to support animation by using 4D $1$-families of neural SDFs.
For this, suppose that the underlying sequence of networks $f_{\theta_1}, \ldots , f_{\theta_m}$ has the \textit{space-time} $\R^3\times \R$ as domain. 
Then, we require the sequence of neural SDFs $f_{\theta_1}(\cdot, t)\rhd \cdots \rhd f_{\theta_m}(\cdot, t)$ to be nested, for each~$t$. See~\cite{novello21neuralAnimation}
for more details on families of neural SDFs. 
Varying $t$ animates $f_{\theta_1}(\cdot, 0)\rhd \cdots \rhd f_{\theta_m}(\cdot, 0)$.

\subsection{Nesting the Neighborhoods}
\label{s-examples}

In this Section, we describe theoretical and practical approaches to create sequences of neural SDFs with nested neighborhoods. The objective is to train a sequence of neural SDFs sorted by inference time and to find small upper-bound thresholds that ensure the nesting condition. 
The supplementary material presents the proofs of the propositions in this Section.

\subsubsection*{\textbf{BACON}}
Since BACON~\cite{lindell2021bacon} is a multiresolution network to represent neural SDFs, its levels of detail are naturally sorted by inference time and can be used to define our neural SDF sequence. Specifically, let $f_{\theta_j}:\R^3\to \R$ be $m$ LODs of a BACON network, and $\epsilon>0$ be a small number. 
Defining $\varepsilon_j:=\abs{f_{\theta_j}\!-\!f_{\theta_{j-1}}}_\infty\!+\epsilon$ results in $\abs{f_{\theta_j}-f_{\theta_{j-1}}}_\infty\!\!<\!\varepsilon_j$;
where $\abs{f}_\infty=\sup\left\{\abs{f(p)};\, p\in \R^3\right\}$.
Then, Proposition~\ref{proposition1} gives the thresholds $\delta_i$ implying that $f_{\theta_j}$ satisfy the nesting condition, i.e. $f_{\theta_1}\rhd \dots\rhd f_{\theta_m}$.
\begin{proposition}\label{proposition1}
Let $f_i$ be $m$ functions satisfying $\abs{f_i-f_{i-1}}_\infty<\varepsilon_i$ for $\varepsilon_i>0$.
This sequence is nested for the thresholds $\elod_i$ defined by $\elod_m=\varepsilon_m$ and $\elod_{i-1}:=\elod_i+\varepsilon_{i}$ for $i=2,\ldots, m$.
\end{proposition}

\subsubsection*{\textbf{MLPs for a single surface}}

MLPs tend to learn lower frequencies first, a phenomenon known as the \textit{spectral bias}~\cite{rahaman2019spectral}. Thus, we propose training MLPs with increasing capacity to represent a single SDF, resulting in a sequence of neural SDFs sorted by inference time and also by detail representation capacity. We use the approach presented in~\cite{novello21diff} to train each neural SDF.

Specifically, let $S$ be a surface, and $f$ be its SDF. Consider $f_{\theta_j}:\R^3\to \R$ to be $m$ MLPs approximating $f$ sorted by capacity, i.e., there are small numbers $\varepsilon_j>0$ such that $\abs{f- f_{\theta_{j}}}_\infty<\varepsilon_j$.
Thus, Proposition~\ref{proposition2} gives the thresholds $\delta_i>0$ which result in $f_{\theta_1}\rhd \dots\rhd f_{\theta_m}$. 
\begin{proposition}\label{proposition2}
Let $f$ be a function and $f_{\theta_i}$ be neural SDFs such that $\abs{f-f_{\theta_i}}_\infty<\varepsilon_i$, where $\varepsilon_i>0$. 
$f_{\theta_i}$ satisfy the nested condition for the thresholds $\elod_i$ defined by $\elod_m=\varepsilon_m+\varepsilon_{m-1}$ and $\elod_{i-1}:=\elod_i+\varepsilon_{i}+\varepsilon_{i-1}$ for $i=2,\ldots, m$.
\end{proposition}

To compute the thresholds $\delta_j$ we need to evaluate the infinity norm $\abs{f}_\infty$ of the underlying network $f$ on its training domain. In practice, we approximate it by $\max\left\{\abs{f(p_i)}\right\}$, where $p_i$ are points sampled from the domain.

\subsubsection*{MLPs for multiple surfaces}
This is a theoretical result that relates surfaces in \textit{level of detail} with the existence of nested sequences of neural SDFs. 

Let $S_1, \ldots, S_m$ be surfaces in level of detail such that $S_j$ deviates no more than a small number $\varepsilon_j>0$ from $S_{j-1}$. That is, $\dist{S_{i-1}}{S_{i}}<\varepsilon_i$, where d is the \textit{Hausdorff distance}. Such sequences of surfaces are considered by~\cite{eck1995multiresolution} in the context of meshes.
Let $\epsilon>0$ be a small number. The \textit{universal approximation theorem} \cite{cybenko1989approximation} states that there are MLPs $f_{\theta_j}$ that deviates no more than $\epsilon$ from the SDFs $f_j$ of $S_j$.
Prop.~\ref{proposition3} provides the thresholds $\delta_i$ that imply $f_{\theta_1}\rhd \dots\rhd f_{\theta_m}$. 
\begin{proposition}\label{proposition3}
Let $S_i$ be $m$ surfaces such that $\dist{S_{i-1}}{S_{i}}\!<\!\varepsilon_i$, with $\varepsilon_i>0$.
For any $\epsilon>0$, there are neural SDFs approximating the SDFs of $S_i$ that are nested for the thresholds defined by $\elod_m=\epsilon$ and $\elod_{i-1}=\elod_i+\varepsilon_{i}+2\epsilon$ for $1<i\leq m$.
\end{proposition}

\section{Multiscale Sphere Tracing}

Let $f_{\theta_1}\rhd \dots\rhd f_{\theta_m}$ be a nested sequence of neural SDFs, $p$ be a point outside $\big[|f_{\theta_1}|\leq\elod_1\big]$, and $v$ be a direction. We define a multiscale ST that approximates the first hit between the ray $\gamma(t)=p+tv$, with $t>0$, and the neural surface $S_{m}$ (see Alg.~\ref{a-sphere-tracing}).
It is based on the fact that to sphere trace $S_{j+1}$ we can first sphere trace $S_j+\delta_j$ using $f_{\theta_{j}}$ 
(see Fig.~\ref{f-sphere_tracing}). Then we continue to iterate in $\big[|f_{\theta_{j}}|\leq\elod_{j}\big]$ using $f_{\theta_{j+1}}$. Lines 3-6 describe the sphere tracing of $S_j+\delta_j$ for $j=1,\ldots, m$ (line 1). 
If $j=m$ we sphere trace the desired surface $S_m$ instead of its neighborhood (line 4).

In the dynamic case, the algorithm operates in $1$-families of nested neural SDFs indexed by the time parameter. 
\begin{algorithm}[hh]
\SetAlgoLined
 \KwInput{A sequence $f_{\theta_1}\rhd \cdots \rhd f_{\theta_{m}}$, position $p$, unit direction $v$, and a threshold $\epsilon>0$.}
 \KwOutput{End point $p$}
 \For{$j=1,\ldots, m$}
 {
    $t=+\infty$\;
    \While{$t>\epsilon$}
    {
        $t \texttt{ = } (j\texttt{==}m)\texttt{ ? } f_{\theta_{j}}(p)\texttt{ : } f_{\theta_{j}}(p)-\elod_j $ \;
        $p\texttt{ = } p+tv$\;
    }
 }
 \caption{Multiscale ST}
 \label{a-sphere-tracing}
\end{algorithm}

If $\gamma\cap S_{m}\neq \emptyset$, we can prove that the multiscale ST approximates the first hit point $q$ between the ray $\gamma$ and $S_{m}$. Indeed, by the nesting condition, if $\gamma\cap (S_{j}+\elod_{j})\neq \emptyset$ implies $\gamma\cap (S_{j-1}+\elod_{j-1})\neq~\emptyset$. 
The classical ST guarantees that iterating $p=p+f_{\theta_{j}}(p)v$ approximates the intersection between $\gamma$ and $S_{j}+\elod_{j}$.
The proof follows by induction.

For the inference of a neural SDF, in line 4 of Alg.~\ref{a-sphere-tracing}, we use the GEMM alg.~\cite{dongarra1990set} for each layer. The final step for rendering is to calculate the normals, given by the gradients of a detailed neural SDF. The next section describes the process.

\section{Neural Normal Mapping}
\label{s-normal-mapping}

\begin{figure}[h]
\centering
    \includegraphics[width=\columnwidth]{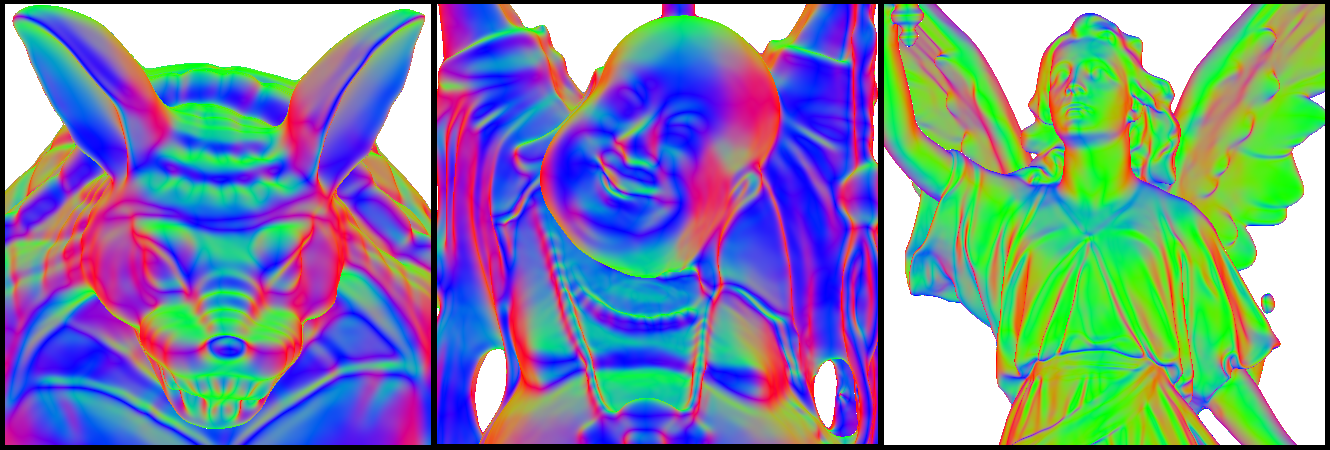}
    \caption{Rendering of the normals calculated using our framework. They are naturally smooth as a consequence of working on the continuous setting.}
    \label{f-normals}
\end{figure}

We propose an analytical computation of normals and a neural normal mapping procedure, which are decoupled from the multiscale ST. Those approaches are continuous, avoiding the need of specific methods to handle discretization, such as \textit{finite differences}. This results in smooth normals, as shown in Figure~\ref{f-normals}.

The idea is to use a finer neural SDF $f_\theta$ to provide the normals of a coarse surface $S$ nested in a 
$\delta$-neighborhood of the zero-level set $S_\theta$, i.e. $S\subset \big[|f_\theta|\leq \delta\big]$. The \textit{neural normal mapping} assigns to each point $p$ in $S$ the normal
$N_\theta(p):=\nabla f_\theta(p).$
This mapping is a restriction of $\nabla f_\theta$ to the surface $S$.
If $\nabla f_\theta$ have no critical points in $\big[|f_\theta|\leq \delta\big]$, each point $p\in S$ can be connected to a point $q\in S_\theta$ by integrating the vector field $-\nabla f_\theta$. 
Thus, the neural normal mapping transports the normal $\nabla f_\theta(q)$ of $S_\theta$, along the resulting path, to the point $p$ by using $\nabla f_\theta(p)$.
If $|\nabla f_\theta|=1$ this path is a straight line and the vector field is constant along~it, implying in $\nabla f_\theta(q)=\nabla f_\theta(p)$ (See~Fig.~\ref{f-overview}).

In practice, we cannot guarantee that $\nabla f_\theta$ has no critical points in $\big[|f_\theta|\leq \delta\big]$. However, in our shading experiments, this is not a problem because the set of critical points of SDFs has a zero measure.

We explore two examples. First, consider $S$ being a triangle mesh. In this case, the neural normal mapping transfers the detailed normals of the level sets of $f_\theta$ to the vertices of $S$. This approach is analogous to the classic normal mapping, which maps detailed normals stored in textures to meshes via parametrizations. However, our method is volumetric, automatic, and does not need such parametrizations. See Fig.~\ref{f-armadillo_ov}.

For the second example, consider $S$ to be the zero-level set of another coarse neural SDF. In this case, we can use the neural normal mapping to avoid the overhead of additional ST iterations. See Fig.~\ref{f-buddha}.
Animated neural SDFs are also supported by mapping the normals of $f_{\theta}(\cdot, t)$ into the animated surface.

\subsection{Analytic Normal Calculation for MLPs}
We present an algorithm based on GEMMs to analytically compute normals for MLP-based neural SDFs, designed for real-time applications. For this, remember that a MLP $f_\theta$ with $n-1$ hidden layers has the following form: 
\begin{equation}\label{e-network-architecture}
    f_\theta(p)\!=\!W_n\circ f_{n-1}\circ f_{n-2}\circ \cdots \circ f_{0}(p)+b_n,
\end{equation}
where $f_{i}(p_i)=\varphi(W_i p_i + b_i)$ is the $i$-layer. The smooth \textit{activation function} $\varphi$ is applied on each coordinate of the linear map $W_i:\R^{N_i}\!\to\!\R^{N_{i+1}}$ translated by $b_i\in\R^{N_{i+1}}$. 

We compute the gradient of $f_{\theta}$ using the \textit{chain~rule}
\begin{align}\label{e-neural_implicit_gradient}
    \grad{f_\theta}(p)\!=\!W_n\cdott \jac f_{n-1} (p_{n-1})\cdott \cdots \cdott \jac f_0(p).
\end{align}
$\jac$ is the \textit{Jacobian} and $p_i:=f_{i-1}\circ \cdots \circ f_{0}(p)$. The Jacobians of $f_i$ applied to the points $p_i$ are given by~\cite{novello21diff}: 
\begin{align}\label{e-jacobian_layers}
    \jac f_{i}(p_i)=W_i\odot \varphi'\big[a_i|\cdots|a_i\big]
\end{align}
where $\odot$ is the \textit{Hadamard} product, and the matrix $\big[a_i|\cdots |a_i\big]$ has $N_i$ copies of the vector $a_i=W_i(p_i)+b_i$. 

The normals of $S_\theta$ are given by $\nabla f_\theta$ which is a sequence of matrix multiplications (Eq~\ref{e-neural_implicit_gradient}).
These multiplications do not fit into a GEMM setting directly since $\jac{f_0}(p)\in \R^{3\times N_1}$.
This is a problem because the GEMM algorithm organizes the input points into a matrix, where its lines correspond to the point coordinates and its columns organize the points and enable parallelism. 
However, we can solve this problem using three GEMMs, one for each normal coordinate. Thus, each GEMM starts with a column of $\jac{f_0}(p)$, eliminating one of the dimensions. The resulting multiplications can be asynchronous since they are completely~independent.

The $j$-coord of $\grad{f_\theta}(p)$ is given by $G_n=W_n\cdott G_{n-1}$, where $G_{n-1}$ is obtained by iterating $G_{i}=\jac f_{i}(p_{i})\cdott G_{i-1}$, with the initial condition $G_0=W_0[j] \odot \varphi'(a_0)$. The vector $W_0[j]$ denotes the $j$-column of the weight matrix $W_0$.

We use a kernel and a GEMM to compute $G_0$ and $G_n$. For $G_{i}$ with $0\!<\!i\!<\!n$, observe~that
\begin{align*}\label{e-Gi}
G_{i}\!=\!\left(W_i\odot \varphi'\left[a_i|\cdots|a_i\right]\right)\cdott G_{i-1}\!=\!(W_i\cdott G_{i-1})\odot \varphi'(a_i).
\end{align*}
The first equality comes from Eq.~\ref{e-jacobian_layers} and the second from a kind of commutative property of the Hadamard product.
The second expression needs fewer computations and is solved using a GEMM followed by a kernel.

Alg.~\ref{a-normal_computation} presents the above gradient computation for a batch of points. 
The input is a matrix $P\in\R^{3\times k}$ with columns storing the $k$ points generated by the GEMM version of Alg.~\ref{a-sphere-tracing}. 
The algorithm outputs a matrix $\grad{f_\theta}(P)\in\R^{3\times k}$, where its $j$-column is the gradient of $f_\theta$ evaluated at $P[j]$.
Lines $2-5$ are responsible for computing $G_0$, lines $6-11$ compute $G_{n-1}$, and line $13$ provides the result $G_n$.
\begin{algorithm}[hh]
\SetAlgoLined
 \KwInput{neural SDF $f_{\theta}$, positions $P$}
 \KwOutput{Gradients $\grad{f_\theta}(P)$}

\For{ $j = 0$ to $2$ (async)} 
{
    using a GEMM: \tcp{Input Layer}
        \Indp $A_0 = W_0\cdott P + b_0$ \\
    \Indm using a kernel: \\
        \Indp $G_0 = W_0[j] \odot \varphi'(A_0)$;\,\,\,
        $P_0 = \varphi(A_0)$ \\
    \Indm
    \tcp{Hidden layers}
    \For{layer $i=1$ to $n-1$}
    {
            using GEMMs: \\
                \Indp $A_i = W_i\cdott P_{i-1} + b_i$;\,\,\,
                $G_i = W_i \cdott G_{i-1}$ \\
            \Indm using a kernel: \\
                \Indp $G_i = G_i \odot \varphi'( A_i)$;\,\,\,
                $P_i = \varphi(A_i)$ \\
    }
    using a GEMM: \tcp{Output layer}
        \Indp $G_n = W_n \cdott G_{n-1}$
}
 \caption{Normal computation}
 \label{a-normal_computation}
\end{algorithm}

\section{Experiments}
We present perceptual/quantitative experiments to evaluate our method. We fix the number of iterations for better control of parallelism. All experiments are conducted on an NVidia Geforce RTX 3090, with all pixels being evaluated (no acceleration structures are used).

We use a simplified notation to refer to the MLP architectures used. For example, ($64, 1) \rhd (256, 3)$ means a neural SDF sequence with a MLP with one $64 \times 64$ matrix (2 hidden layers with 64 neurons), and a MLP with three $256 \times 256$ matrices (4 hidden layers with 256 neurons). Another example: $(64, 1)$ is a single MLP.

\begin{figure}[h]
\centering
    \includegraphics[width=0.85\columnwidth]{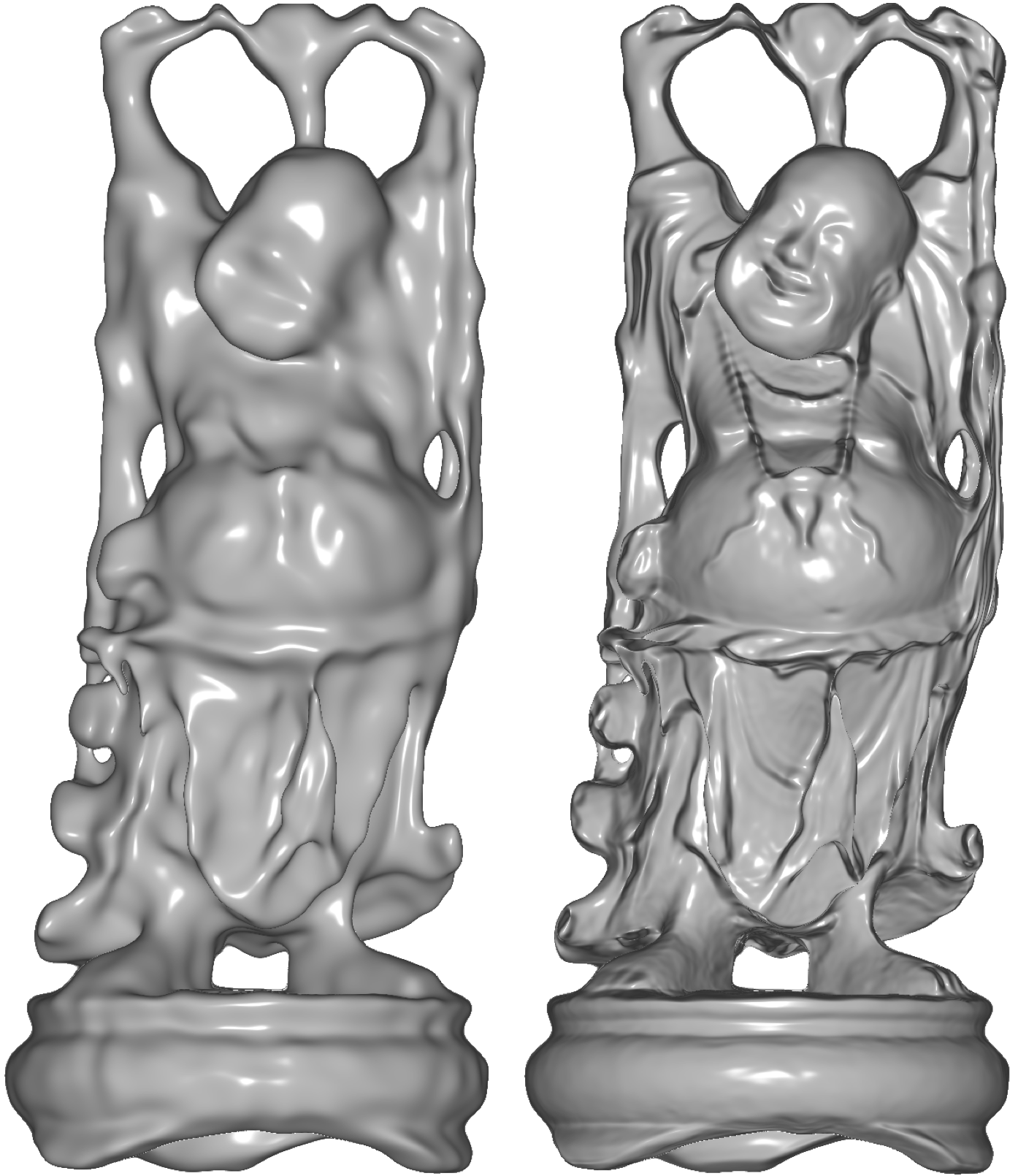}
    \caption{Neural Normal Mapping into a neural SDF surface. On the left, a $\boldsymbol{(64, 1)}$ neural SDF without normals mapped. On the right, a neural normal mapping of the $\boldsymbol{(256, 3)}$ neural SDF into the $\boldsymbol{(64, 1)}$.}\label{f-buddha}
\end{figure}

\begin{figure}[h]
\centering
    \includegraphics[width=0.85\columnwidth]{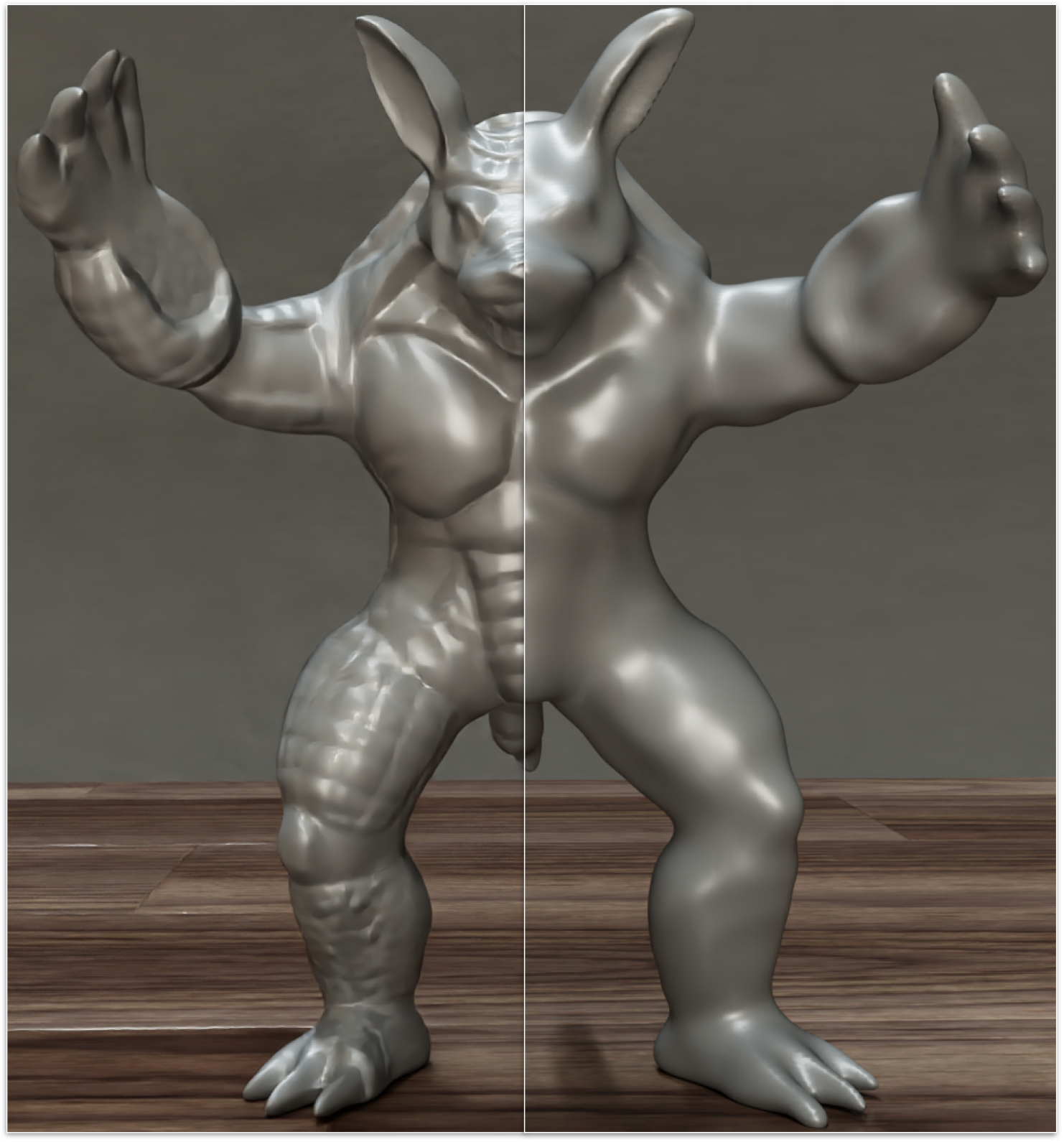}
    \caption{Neural normal mapping into a triangle mesh surface. On the left side, the normals of a neural SDF $(256,3)$ are mapped into a coarse surface. On the right side, the original normals are used. The underlying surface is the same for both cases (marching cubes of a $(64,1)$ neural SDF). The MSE is $0.00262$ for the coarse case and $0.00087$ for the normal mapping, an improvement of $3$x. The baseline is the marching cubes of the $(256,3)$ neural SDF.}
    \label{f-armadillo_ov}
\end{figure}

\begin{figure*}
\centering
    \includegraphics[width=0.88\textwidth]{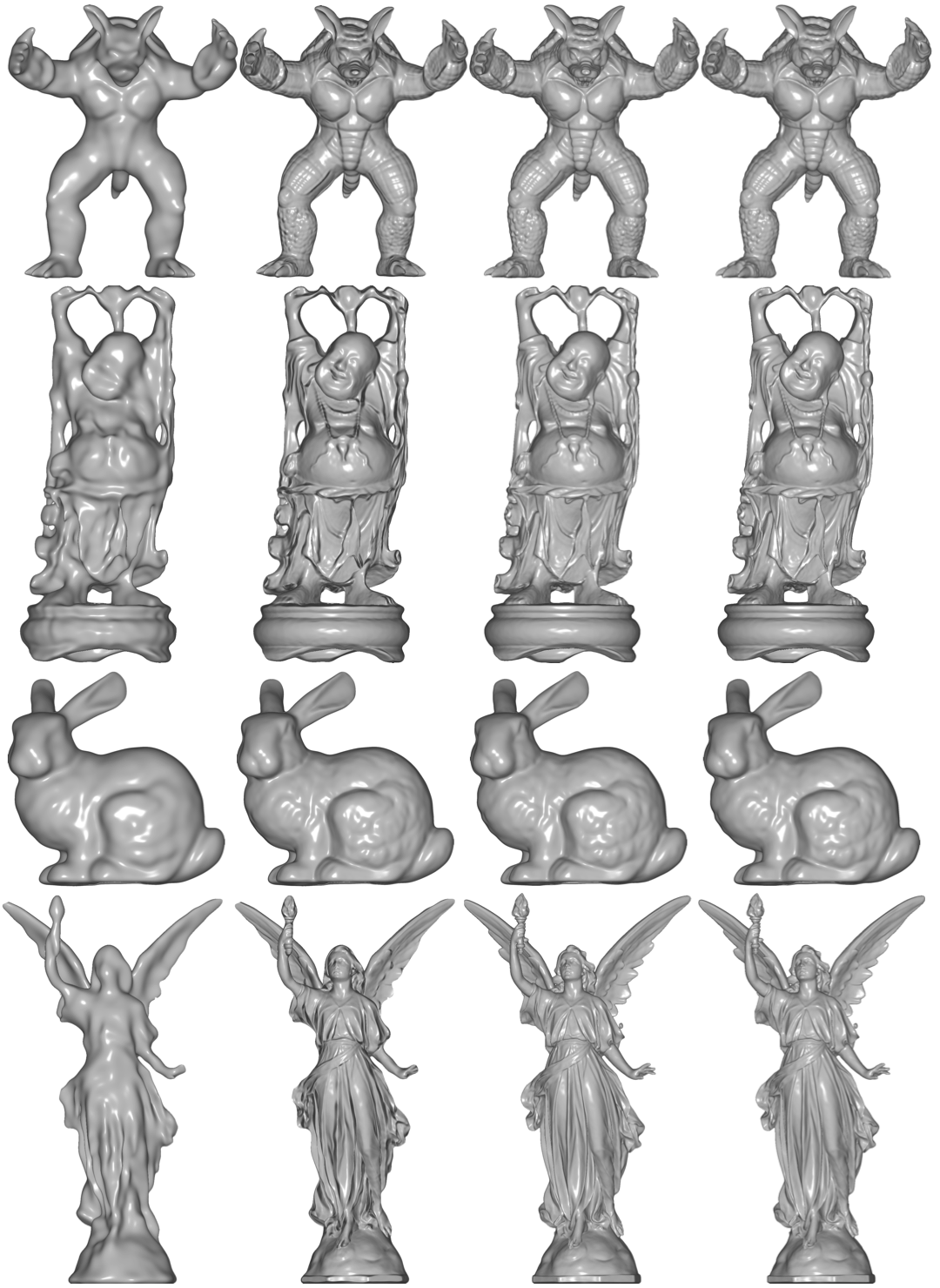}
    \caption{Comparison between our method and the SIREN baseline. The columns represent different configurations. From left to right: $\boldsymbol{(64, 1)}$, $\boldsymbol{(64, 1) \rhd (256, 1)}$ (Bunny and Dragon) and $\boldsymbol{(64, 1) \rhd (256, 2)}$ (Happy Buddha and Lucy), $\boldsymbol{(64, 1) \rhd (256, 3)}$, and the baseline $\boldsymbol{(256, 3)}$. Notice how the second column is already similar to the baseline. The third column adds more detail.}\label{f-rendering}
\end{figure*}

First, we discuss the two applications of neural normal mapping. Regarding image quality/perception, Figs.~\ref{f-buddha} and \ref{f-armadillo_ov} show the case where the coarse surface is the zero-level of a neural SDF and when it is a triangle mesh, respectively. An overall evaluation of the framework is presented in Fig.~\ref{f-rendering}. 
In all cases, normal mapping increases fidelity considerably.

The quantitative results corroborate the statement above. Table~\ref{t-ablation} shows performance metrics regarding time, memory, and mean square error (MSE) measured in a Python renderer. An important remark is that the neural normal mapping increases fidelity (up to $30\%$ MSE improvement in comparison with the Armadillo coarse case) and considerably accelerates the rendering as well (up to 6x improvement in comparison with the Armadillo baseline).

\begin{table}[h]
\caption{Evaluation of our method considering two SIRENs in a Python renderer. Iters represent the number of iterations used in each one (0 means no iteration and thus pure neural normal mapping). Time is in seconds and memory is in KB. MSE is the mean square error compared with the baseline (in italic). We emphasize in bold how the neural normal mapping has minimal time impact and how increasing iterations on the second SDF improves MSE.}
\label{t-ablation}
\small
\begin{tabular}{l|llllc}
\cline{2-6}
                                           & \textbf{Nets}    & \textbf{Iters} & \textbf{Time} & \textbf{Mem} & \textbf{MSE}     \\ \hline
\parbox[t]{2mm}{\multirow{6}{*}{\rotatebox[origin=c]{90}{\textbf{Armadillo}}}}
 & \textit{(256,3)} & \textit{40}    & \textit{2.442}   & \textit{777}     & \textit{-} \\
\multicolumn{1}{c|}{}                                    & $(64,1)$           & 40             & 0.298            & 18               & 0.00588          \\
\multicolumn{1}{c|}{}                                    & $(64,1)\rhd (256,3)$  & 40, 0          & \textbf{0.409}   & 795              & 0.00452          \\
\multicolumn{1}{c|}{}                                    & $(256,3)$          & 15             & 0.936            & 777              & 0.01237          \\
\multicolumn{1}{c|}{}                                    &$(64,1)\rhd (256,3)$  & 30,10          & 0.895            & 795              & 0.00746          \\
\multicolumn{1}{c|}{}                                    & $(64,1)\rhd (256,3)$  & 30,30          & 1.934            & 795              & \textbf{0.00057} \\ \hline
\parbox[t]{2mm}{\multirow{6}{*}{\rotatebox[origin=c]{90}{\textbf{Buddha}}}}                        & \textit{(256,3)} & \textit{40}    & \textit{2.228}   & \textit{777}     & \textit{-} \\
                                                         & (64,1)           & 40             & 0.299            & 18               & 0.00485          \\
                                                         & $(64,1)\rhd (256,3)$  & 40, 0          & \textbf{0.413}   & 795              & 0.00441          \\
                                                         & (256,3)          & 15             & 0.928            & 777              & 0.00589          \\
                                                         & $(64,1)\rhd (256,3)$  & 30,10          & 0.893            & 795              & 0.00355          \\
                                                         & $(64,1)\rhd (256,3)$  & 30,30          & 1.945            & 795              & \textbf{0.00048} \\ \hline
\parbox[t]{2mm}{\multirow{6}{*}{\rotatebox[origin=c]{90}{\textbf{Bunny}}}}                        & \textit{(256,3)}          & \textit{40}             & \textit{2.237}            & \textit{777}              & \textit{-}          \\
                                                         & (64,1)           & 40             & 0.287            & 18               & 0.00229          \\
                                                         & $(64,1)\rhd (256,3)$  & 40, 0          & \textbf{0.403}   & 795              & 0.00191          \\
                                                         & (256,3)          & 15             & 0.928            & 777              & 0.00793          \\
                                                         & $(64,1)\rhd (256,3)$  & 30,10          & 0.886            & 795              & 0.00417          \\
                                                         & $(64,1)\rhd (256,3)$  & 30,30          & 1.920            & 795              & \textbf{0.00065} \\ \hline
\parbox[t]{2mm}{\multirow{6}{*}{\rotatebox[origin=c]{90}{\textbf{Lucy}}}}                     & \textit{(256,3)}          & \textit{40}             & \textit{2.239}            & \textit{777}              & \textit{-}          \\
                                                         & (64,1)           & 40             & 0.312            & 18               & 0.00518          \\
                                                         & $(64,1)\rhd (256,3)$  & 40, 0          & \textbf{0.420}   & 795              & 0.00470          \\
                                                         & (256,3)          & 15             & 0.941            & 777              & 0.00280          \\
                                                         & $(64,1)\rhd (256,3)$  & 30,10          & 0.927            & 795              & 0.00363          \\
                                                         & $(64,1)\rhd (256,3)$  & 30,30          & 1.977            & 795              & \textbf{0.00024} \\ \hline
\end{tabular}
\end{table}

\begin{figure}[h]
\centering
    \includegraphics[width=\columnwidth]{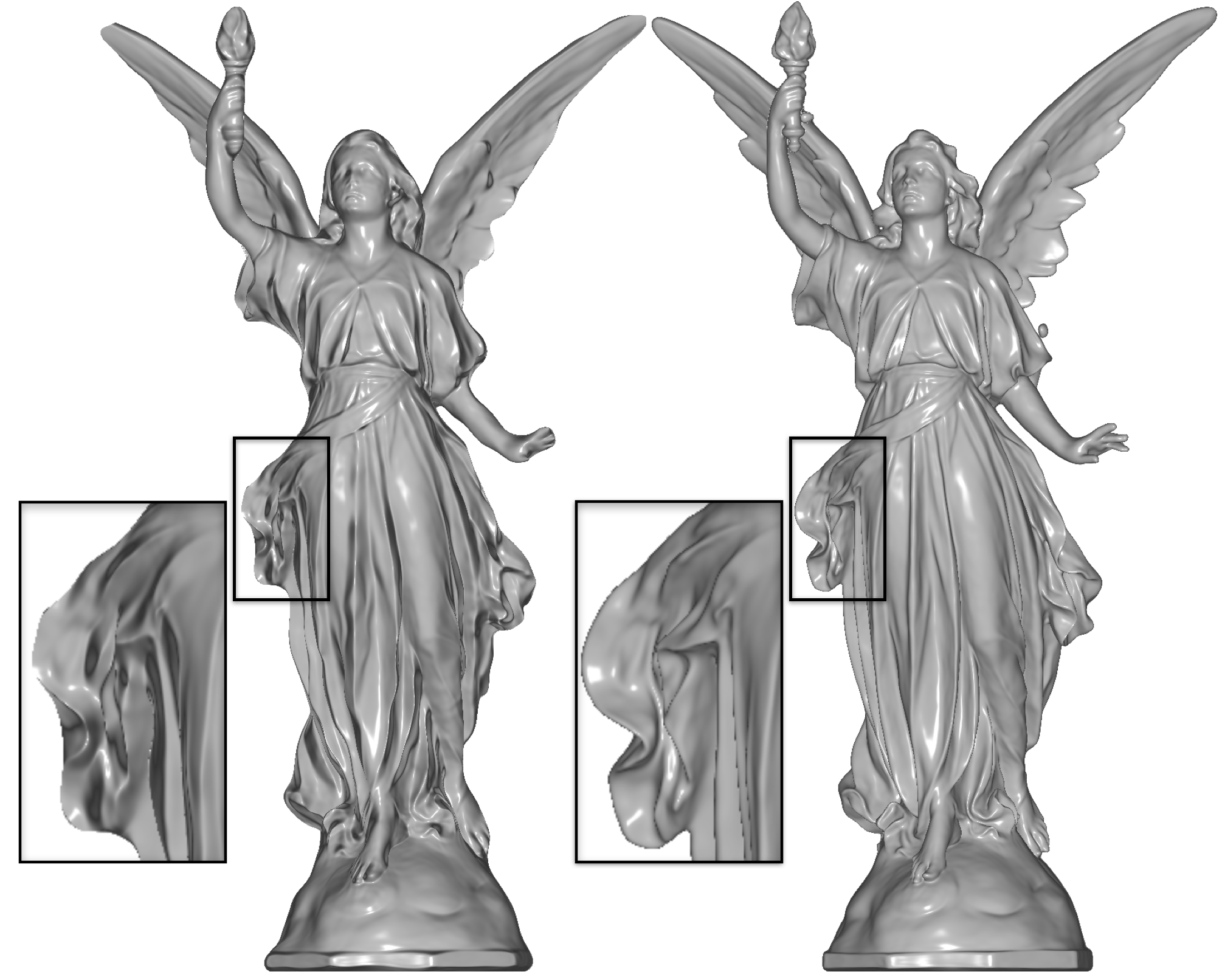}
    \caption{Silhouette evaluation. Left: $\boldsymbol{(64, 1)\rhd (256,3)}$, right: $\boldsymbol{(64, 1)\rhd (256,2)\rhd (256,3)}$. Notice how the silhouettes improve with the additional $\boldsymbol{(256,2)}$ level.}\label{f-silhouette}
\end{figure}

The result may be improved using the multiscale ST, as shown in Figure~\ref{f-silhouette}. Adding ST iterations using a neural SDF with a better approximation of the surface improves the silhouette. This is aligned with the results presented in Table~\ref{t-ablation}. The last two rows of each example show cases with iterations in the second neural SDF, with considerable improvement in the MSE (up to 20x improvement in comparison with the pure neural normal mapping for Lucy). 

Table~\ref{t-bacon} shows a comparison with BACON in a Python renderer. We use a BACON network with 8 layers, with output in layers 2 and 6. For fairness, we use the same layers as neural SDFs for our method. The interpretation is analogous to the one for Table~\ref{t-ablation}. To compare with the SIREN case, please refer to the numbers in that table.

\begin{table}[]
\caption{Comparison with BACON~\cite{lindell2021bacon} in a Python renderer. We use a BACON network with 8 layers, with output in layers 2 and 6 For fairness, we use the same layers as SDFs for our method. This Table follows the same structure of Table~\ref{t-ablation}.}
\label{t-bacon}
\small
\begin{tabular}{l|llrlc}
\cline{2-6}
                                    & \textbf{Nets}         & \textbf{Iters} & \textbf{Time}  & \textbf{Mem}     & \textbf{MSE}     \\ \hline
\parbox[t]{2mm}{\multirow{5}{*}{\rotatebox[origin=c]{90}{\textbf{Armadillo}}}} & \textit{(256,6)}    & \textit{100}   & \textit{10.067} & \textit{2151} & \textit{-} \\
                                    & $(256,2)$             & 100            & 4.829          & 2151          & 0.00473          \\
                                    & $(256,2)\rhd(256, 6)$ & 100,0          & \textbf{4.945} & 2151          & 0.00309          \\
                                    & $(256,2)\rhd(256, 6)$ & 50,30          & 5.526          & 2151          & 0.00061          \\
                                    & $(256,2)\rhd(256, 6)$ & 50,50          & 7.438          & 2151          & \textbf{0.00040} \\ \hline
\parbox[t]{2mm}{\multirow{5}{*}{\rotatebox[origin=c]{90}{\textbf{Buddha}}}}     & \textit{(256,6)}    & \textit{100}   & \textit{9.851} & \textit{2151} & \textit{-} \\
                                    & $(256,2)$             & 100            & 4.836          & 2151          & 0.00455          \\
                                    & $(256,2)\rhd(256, 6)$ & 100,0          & \textbf{4.946} & 2151          & 0.00284          \\
                                    & $(256,2)\rhd(256, 6)$ & 50,30          & 5.520          & 2151          & 0.00086          \\
                                    & $(256,2)\rhd(256, 6)$ & 50,50          & 7.450          & 2151          & \textbf{0.00077}          \\ \hline
\parbox[t]{2mm}{\multirow{5}{*}{\rotatebox[origin=c]{90}{\textbf{Bunny}}}}      & \textit{(256,6)}    & \textit{100}   & \textit{9.861} & \textit{2151} & \textit{-} \\
                                    & $(256,2)$             & 100            & 4.835          & 2151          & 0.00458          \\
                                    & $(256,2)\rhd(256, 6)$ & 100,0          & \textbf{4.952} & 2151          & 0.00260          \\
                                    & $(256,2)\rhd(256, 6)$ & 50,30          & 5.524          & 2151          & 0.00025          \\
                                    & $(256,2)\rhd(256, 6)$ & 50,50          & 7.455          & 2151          & \textbf{0.00013}          \\ \hline
\parbox[t]{2mm}{\multirow{5}{*}{\rotatebox[origin=c]{90}{\textbf{Lucy}}}}       & \textit{(256,6)}    & \textit{100}   & \textit{9.871} & \textit{2151} & \textit{-} \\
                                    & $(256,2)$             & 100            & 4.852          & 2151          & 0.00400          \\
                                    & $(256,2)\rhd(256, 6)$ & 100,0          & \textbf{4.968} & 2151          & 0.00207          \\
                                    & $(256,2)\rhd(256, 6)$ & 50,30          & 5.559          & 2151          & 0.00023          \\
                                    & $(256,2)\rhd(256, 6)$ & 50,50          & 7.488          & 2151          & \textbf{0.00018}          \\ \hline
\end{tabular}
\end{table}

We evaluated a GPU version implemented in a CUDA renderer, using neural normal mapping, multiscale ST, and the analytical normal calculation (with GEMM implemented using CUTLASS). Table~\ref{t-3d} shows the results. Notice that the framework achieves real-time performance and that using neural normal mapping and multiscale ST improves performance considerably.

\begin{table}[h]
  \caption{Real-time evaluation using SIRENs with our analytical normal calculation in a CUDA renderer. The number of iterations is 20 for the first neural SDF and 5 for the second neural SDF in the last row. The last network is used for neural normal mapping only. All pixels are evaluated. Images are~$512~\times~512$. Memory is in KB. Notice that all cases using multiscale ST and neural normal mapping result in speedups.}
  \label{t-3d}
  \small
  \begin{tabular}{lrrr}
    \hline
    Model&FPS&Speedup&Mem\\
    \hline
    $$(256,3)$$ & 19.8 & 1.0 & 777 \\
    $(64, 1)$ & \textbf{128.8} & \textbf{6.5} & 18\\
    $(64, 1) \rhd (256, 1)$ & \textbf{73.1} & \textbf{3.7} & 281\\
    $(64, 1) \rhd (256, 2)$ & \textbf{53.0} & \textbf{2.7} & 538\\
    $(64, 1) \rhd (256, 3)$ & \textbf{41.6} & \textbf{2.1} & 795\\
    $(64, 1) \rhd (256, 1) \rhd (256, 3)$ & \textbf{39.1} & \textbf{2.0} & 1058\\
  \hline
\end{tabular}
\end{table}

\begin{figure}[h]
\centering
    \includegraphics[width=\columnwidth]{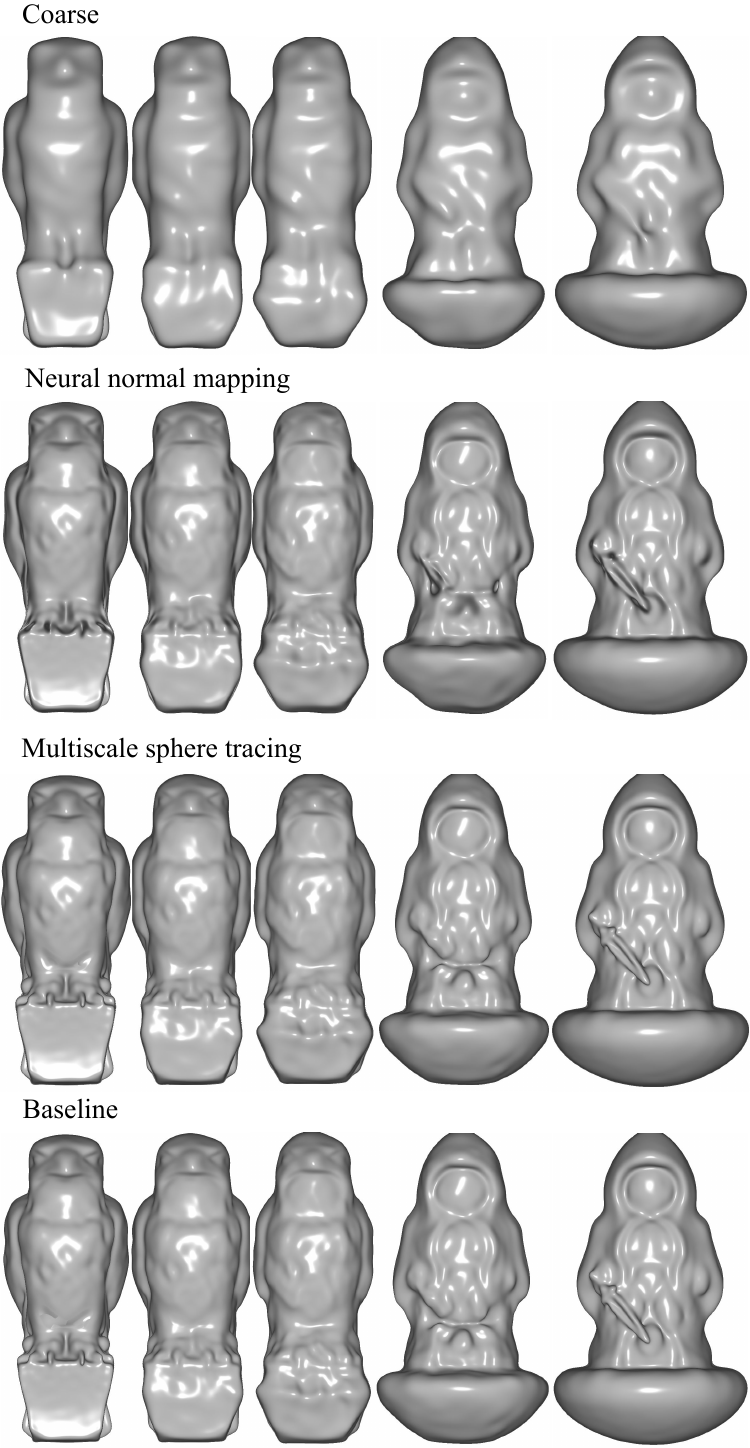}
    \caption{Interpolation between the Falcon and Witch models. From top to bottom: $\boldsymbol{(64, 1)}$, $\boldsymbol{(64, 1) \rhd (128,2)}$, and the baseline $\boldsymbol{(128, 2)}$. Notice how the normal mapping in the second line increases~fidelity. See the video in the supplementary material.}\label{f-i4d}
\end{figure}

Finally, Fig.~\ref{f-i4d} shows an evaluation of an animated neural SDF representing the interpolation of the Falcon and Witch models. The baseline neural SDF is $(128,2)$ and the coarse is $(64,1)$. The example normal mapping case runs in the CUDA renderer at 73FPS. See the supplementary video for the full animation.

\section{Conclusion}

We presented a novel approach to render neural SDFs based on three decoupled algorithms: the multiscale ST, the neural normal mapping, and the analytic normal calculation for MLPs. Those algorithms support animated 3D models and do not need spatial data structures to work. Neural normal mapping can also be used on contexts outside neural SDFs, enabling smooth normal fetching for discretized representations such as meshes as well.

This work opens paths for several future work options. For example, exploring the mapping for other attributes could be interesting. Possible candidates include material properties, BRDFs, textures, and hypertextures.

Another path to explore is performance. Improvements can be done for further optimization. For example, using fully fused GEMMs may decrease the overhead of GEMM setup~\cite{tiny-cuda-nn}. The framework may also be adapted for acceleration structures and ray tracing engines such as OptiX.

{\small
\setlength{\bibsep}{0pt}
\bibliographystyle{abbrvnat}

\bibliography{sample-base}
}

\end{document}